\documentclass{amsart}
\usepackage{graphicx}
\usepackage{amssymb, amsmath}
\usepackage{amsfonts}
\usepackage{amssymb}
\usepackage{epsfig}
\usepackage{float}

\newtheorem{theorem}{Theorem}

\newtheorem{definition}[theorem]{Definition}

\newtheorem{lemma}[theorem]{Lemma}

\newtheorem{remark}[theorem]{Remark}

\newcommand{\divv}{\text{\rm div}}
{\bf }

\newcommand{\cL}{\mathcal L}
\newcommand{\C}{\mathbb C}

\newcommand{\R}{\mathbb R}

\newcommand{\dist}{\text{\rm dist}}
\def\be{\begin{equation}}
\def\ee{\end{equation}}
\def\la{\lambda}

\def\R{\mathbb R}

\def\tilde{\widetilde}

\numberwithin{equation}{section}
\numberwithin{theorem}{section}
\numberwithin{figure}{section}

\begin{document}
\bibliographystyle{siam}

\title[Stratified Rotating Boussinesq Equations]{Stratified Rotating Boussinesq Equations in Geophysical Fluid Dynamics: Dynamic Bifurcation and Periodic Solutions}

\author[C. Hsia]{Chun-Hsiung Hsia}
\address[CH]{Department of Mathematics, Statistics,
and Computer Science, The University of Illinois at Chicago, Chicago, IL 60607}
\email{willhsia@uic.edu}

\author[T. Ma]{Tian Ma}
\address[TM]{Department of Mathematics, Sichuan University,
Chengdu, P. R. China}

\author[S. Wang]{Shouhong Wang}
\address[SW]{Department of Mathematics,
Indiana University, Bloomington, IN 47405}
\email{showang@indiana.edu}
\date{\today}

\thanks{The work was supported in part by the
Office of Naval Research,  by the National Science Foundation,
and by the National Science Foundation
of China.}


\begin{abstract}
The main objective of this article is to study the dynamics of the
stratified rotating Boussinesq equations, which are a basic model in
geophysical fluid dynamics.  First, for the case where the Prandtl
number is greater than one, a complete stability and bifurcation
analysis near the first critical Rayleigh number is carried out.
Second, for the case where the Prandtl number is smaller than one,
the onset of the Hopf bifurcation near the first critical Rayleigh
number is established, leading to the existence of nontrivial
periodic solutions. The analysis is based on a newly developed bifurcation and stability theory for nonlinear dynamical systems (both finite and infinite dimensional) by two of the authors \cite{b-book}.
\end{abstract}

\maketitle
\section{Introduction}
The phenomena of the atmosphere and ocean are extremely rich
in its organization and complexity, and a lot of them cannot be produced
by experiments. These phenomena involve  a broad range of
temporal and spatial scales. As we know, both
the atmospheric and oceanic flows are flows
under the rotation of the earth. In fact, fast
rotation and small aspect ratio are two
main characteristics of the large scale atmospheric
and oceanic flows. The small aspect ratio
characteristic leads to the primitive
equations, and the fast rotation leads
to the quasi-geostrophic equations. These
are fundamental equations in the study of
atmospheric and oceanic flows; see  Ghil
and Childress \cite{GC}, Lions, Temam and
Wang \cite{LTWa, LTWb}, and Pedlosky \cite{Pb}.
Furthermore, convection occurs in many regimes of
the atmospheric  and oceanic flows.

A key problem in the study of climate dynamics and in geophysical fluid dynamics is to
understand and predict the periodic, quasi-periodic, aperiodic, and fully turbulent characteristics of large-scale atmospheric and oceanic flows. Stability/bifurcation theory enables one to  determine how different flow regimes appear and disappear as control  parameters, such as the Reynolds number, vary. It, therefore, provides one with a powerful tool to explore the theoretical capability in the predictability problem. Most studies so far have only considered systems of ordinary differential equations (ODEs) that are obtained by projecting the PDEs onto a finite-dimensional solution space, either
by finite differencing or by truncating a Fourier expansion (see Ghil and Childress \cite{GC} and further references there). These  were pioneered by Lorenz \cite{La, Lb}, Stommel \cite{S}, and Veronis \cite{Va, Vb} among others, who explored the bifurcation
structure of low-order models of atmospheric and oceanic flows. More recently, 
pseudo-arclength continuation methods have been applied
to atmospheric (Legras and Ghil \cite{LG}) and oceanic (Speich et al. \cite{SDG} and Dijkstra \cite{D}) models with increasing horizontal resolution. These numerical bifurcation studies have produced so far fairly reliable results for two classes of geophysical flows: (i) atmospheric flows in a periodic mid-latitude channel, in the presence of bottom topography and a forcing jet; and (ii) oceanic
flows in a rectangular mid-latitude basin, subject to
wind stress on its upper surface; see among others Charney and DeVore \cite{CD},
Pedlosky \cite{Pa}, Legras and Ghil \cite{LG} and Jin and Ghil \cite{JG}  for saddle-node and Hopf bifurcations in the the atmospheric channel, and
\cite{BM, CI, IS, JJG, MB, SDG}  for saddle-node, pitchfork or Hopf
in the oceanic basin.

The main objective of this article is to conduct bifurcation and stability analysis  for
the original partial differential equations (PDEs) that govern geophysical flows. This  approach should allow us to overcome some of the inherent limitations of the numerical bifurcation results that dominate the climate dynamics literature up to this point,  and to capture the essential dynamics of the governing PDE systems.

The present article addresses  the stability and transitions of basic flows
for the stratified rotating Boussinesq equations. These equations are fundamental equations in the geophysical fluid dynamics; see among others  Pedlosky \cite{Pb}.
We obtain two main results in this article. The first is to conduct a rigorous and complete
bifurcation and stability analysis near the first eigenvalue of the linearized problem. The second is  the onset of the Hopf bifurcation, leading to the existence of periodic solutions of the model.

The detailed analysis is carried out in two steps. The first  is a detailed study of the
eigenvalue problem for the linearized problem around the basic state.
In comparison to the classical B\'enard convection problem,
the linearized problem here is  non-selfadjoint, leading to much more
complicated spectrum, and more complicated dynamics. We derive
 in particular two critical
 Rayleigh numbers $R_{c_1}$ and $R_{c_2}$.  Here $R_{c_1}$ is
 the first critical Rayleigh number for the case where
 the Prandtl number is greater than one, and $R_{c_2}$ is
 the first critical Rayleigh number for the case where
 the Prandtl number is less than one. Moreover, $R_{c_1}$
 leads to the onset of the steady state
 bifurcation while $R_{c_2}$  leads
  to the onset of the Hopf bifurcation. Both
  parameters are explicitly given
  in terms of the physical parameters.
The crucial issues here include  1) a complete
understanding of the spectrum, 2) identification
of the critical Rayleigh numbers, and
most importantly 3) the verification of
the Principle of Exchange of Stabilities
 near these critical Rayleigh numbers.

The second step is to conduct a rigorous nonlinear analysis to
derive the bifurcations at both the critical Rayleigh numbers based
 on the classical Hopf bifurcation theory and a
 newly developed dynamic bifurcation theory by two of the authors. This new dynamic bifurcation theory  is centered at a new notion of bifurcation,
 called attractor bifurcation for dynamical systems,
  both finite dimensional and infinite dimensional, together with
  new strategies for the Lyapunov-Schmidt reduction and the center manifold
reduction procedures. The bifurcation theory has been applied to
various problems from science and engineering, including, in particular,
the Kuramoto-Sivashinshy equation, the Cahn-Hillard equation,
the Ginzburg-Landau equation, Reaction-Diffusion equations
in Biology and Chemistry, and the B\'enard convection problem, the Taylor problem;
see \cite{b-book, mw-db1} and the references therein.

We remark that the non-selfadjointness of the linearized
problem gives rises the onset of the Hopf Bifurcation. We prove that
the Hopf bifurcation appears at the Rayleigh number $R_{c_2}$. As
mentioned earlier, the understanding and prediction of of the the
periodic, quasi-periodic, aperiodic, and fully turbulent
characteristics of large-scale atmospheric and oceanic flows are key
issues in the study of climate dynamics and in geophysical fluid
dynamics. It is hoped that the study carried out in this article
will provide some insights in these important issues.

Also, we would like to mention that rigorous proof of  the existence of periodic solutions for a fluid system is a normally a very difficult task from the mathematical point of view. For instance, with a highly involved analysis,  Chen et al. \cite{cgsw} proved the existence of a Hopf bifurcation in an idealized Fourier space.

The paper is organized as follows. Section 2 gives the basic setting of the problem.
Section 3 states the main results. The proofs of the main results occupies the
remaining part of the paper: Section 4 recapitulates the essentials of the attractor
bifurcation theory, Section 5 is on the eigenanalysis, and Section 6 is on the central manifold reduction and the completion of the proofs.

\section{Stratified Rotating Boussinesq Equations in Geophysical Fluid Dynamics}
The stratified rotating  Boussinesq equations are basic equations in the geophysical fluid dynamics, and their non-dimensional form is given by
\be
\label{b1}
\left\{
\begin{aligned}
& \frac{\partial U}{\partial t}=\sigma( \Delta U-\nabla p)
  +\sigma R T e - \frac{1}{Ro} e \times U-(U \cdot \nabla)U,\\
& \frac{\partial T}{\partial t}=\Delta T +w-(U \cdot \nabla)T,\\
& \divv  U = 0,
\end{aligned}
\right. \ee 
for $(x,y,z)$ in the non-dimensional domain
$\Omega=\mathbb R^{2}\times (0,1)$, where $U=(u,v,w)$ is the
velocity fields, $e=(0,0,1)$ is the unit vector in the
$z$-direction, $\sigma$ is the Prandtl number, $R$ is the thermal
Rayleigh number, $Ro$ is the Rossby number, $T$ is the temperature
function and $p$ is the pressure function. We refer the interested
readers to Pedlosky \cite{Pb}, Lions, Temam and Wang \cite{LTWb} for
the derivation of this model and the related parameters. In
particular, the term $\frac{1}{Ro} e \times U$ represents the
Coriolis force, the $w$ term in the temperature equation is derived
using the stratification, and  the definition of the Rayleigh number
$R$ as follows:
\begin{equation}
\label{eq1.1}
R = \frac{g\alpha\beta}{\kappa\nu}\, h^4.
\end{equation}

We consider the periodic boundary condition in the $x$ and $y$ directions
\begin{align}
\label{b2}
 (U,T)(x,y,z,t)& =(U,T)(x+2j\pi/\alpha_1,y,z,t) \\
&   =(U,T)(x,y+2k\pi/\alpha_2,z,t), \nonumber
\end{align}
for any $j,k\in \mathbb Z$. At the top and bottom boundaries, we impose
 the free-free boundary conditions:
\be
\label{b3}
(T, w)=0, \quad \frac{\partial u}{\partial z}=0,
\quad \frac{\partial v}{\partial z}=0, \quad \text{at} \quad z=0,1.
\ee

It is natural to put the constraint
\be
\label{b4}
 \int_{\Omega}udxdydz= \int_{\Omega}vdxdydz =0.
\ee

The initial value conditions are given by
\be
\label{b5}
(U,T)=(\tilde{U},\tilde{T}) \quad  \text{at}\quad t=0.
\ee
Let
\begin{align*}
 H=& \{(U,T)\in L^{2}(\Omega)^{4}\mid \divv \, U=0, w\mid_{z=0,1}=0,
      (u,v) \text{ satisfies}\,\, (\ref{b2})\,\,
    \text{and}\,\, (\ref{b4})\},\\
 H_1=& \{(U,T)\in H^{2}(\Omega)^{4}\cap H \,|\,(U,T)\,\text{ satisfies }
  \, (\ref{b2})-(\ref{b4})  \},\\
 \tilde{H}=& \{(U,T)\in H \mid (u,v,w,T)(-x,-y,z)=(-u,-v,w,T)(x,y,z)\},
\\
 \tilde{H}_1=& H_1 \cap \tilde{H}.
\end{align*}
Let
$L_{R}=-A-B_{R}: H_{1} \to H$ (resp., $\tilde{H}_1$ $\to$  $\tilde{H}$)
 and $G: H_{1}$ $\to$ $H$ (resp., $\tilde{H}_1$ $\to$  $\tilde{H}$)
be defined by
\begin{align*}
& A\psi = ( -P [\sigma \Delta U-\frac{1}{Ro} e \times U], -\Delta T), \\
& B_{R} \psi =  (-P [\sigma R T e],-w ),\\
& G (\psi) = G(\psi, \psi),
\end{align*}
for any $\psi=(U, T) \in H_1$ (resp., $\tilde{H}_1$),
where
\begin{align*}
& G(\psi_1 , \psi_2)= (-P[(U_1 \cdot \nabla)U_2], -(U_1 \cdot \nabla)T_2),
\end{align*}
for any $\psi_1=(U_1,T_1)$, $\psi_2 = (U_2, T_2) \in H_1$.
 Here $P$ is the Leray projection to
$L^2$ fields, and for a detailed account of the function spaces;
see among many others \cite{temam}.
\begin{remark}
\label{rmb1}
{\rm
Note that $\tilde{H}_1$ and $\tilde{H}$ are invariant
under the bilinear operator  $G$ in the sense that
$$
G(\psi_1, \psi_2) \in \tilde{H}, \qquad
\text{for} \qquad \psi_1, \psi_2 \in \tilde{H}_1.
$$
Hence,  $\tilde{H}_1$ and $\tilde{H}$ are invariant
under the operator $L_R + G$.
}
\end{remark}

Then the Boussinesq equations (\ref{b1})-(\ref{b4}) can be written
in the following operator form
\begin{equation}\label{b6}
\frac{d\psi}{dt} = L_{R} \psi + G(\psi), \qquad \psi=(U,T).
\end{equation}

\section{Main Results}

\subsection{Definition of attractor bifurcation}
 To state the main theorems of this article,
we proceed with the definition of attractor bifurcation, first introduced by
T. Ma and S. Wang in \cite{b-book,mw-db1}.

Let  $H$ and  $H_1$ be two Hilbert spaces,
and $H_1 \hookrightarrow H$ be a dense and compact inclusion.
We consider the following
nonlinear evolution equations
\be
\label{c1}
\left\{
\begin{aligned}
& \frac{du}{dt} = L_\lambda u +G(u,\lambda), \\
& u(0) = u_0,
\end{aligned}
\right.
\ee
where $u: [0, \infty) \to H$  is the unknown function, $\lambda \in
\mathbb R$  is the  system  parameter, and
$L_\lambda:H_1\to H$ are parameterized linear completely
continuous fields depending continuously on $\lambda\in \R^1$, which
satisfy
\begin{equation}
\label{c2}
\left\{\begin{aligned}
& -L_\lambda = A + B_\lambda && \text{a sectorial operator}, \\
& A:H_1 \to H && \text{a linear homeomorphism}, \\
& B_\lambda :H_1\to H && \text{parameterized linear compact
operators.}
\end{aligned}\right.
\end{equation}
It is easy to see \cite{henry} that $L_\lambda$
generates an analytic semi-group $\{e^{tL_\lambda}\}_{t\ge 0}$.
Then we can define  fractional power operators $(-L_\lambda)^{\mu}$ for any
$0\le \mu \le 1$ with domain $H_\mu = D((-L_\lambda)^{\mu})$ such that
$H_{\mu_1} \subset H_{\mu_2}$ if $\mu_1 > \mu_2$, and $H_0=H$.

Furthermore, we assume that the nonlinear terms
$G(\cdot, \lambda):H_\mu \to H$ for some $1> \mu \ge 0$
are a family of parameterized $C^r$
bounded operators ($r\ge 1$) continuously depending on the parameter
$\lambda\in \R^1$, such that
\begin{equation}
\label{c3}
 G(u,\lambda) = o(\|u\|_{H_\mu}), \quad \forall\,\, \lambda\in \R^1.
\end{equation}

In this paper, we are interested in the sectorial operator
$-L_\lambda = A +B_\lambda$ such
 that there exist an eigenvalue sequence $\{\rho_k\}
\subset \C^1$ and an eigenvector sequence $\{e_k, h_k\}\subset
H_1$ of $A$:
\begin{equation}
\label{c4}
\left\{\begin{aligned}
& Az_k = \rho_kz_k,  \qquad z_k=e_k + i h_k, \\
& \text{Re} \rho_k\to \infty \,\,(k\to\infty), \\
& |\text{Im} \rho_k / (a + \text{Re} \rho_k) | \le c,
\end{aligned}\right.
\end{equation}
for some $a, c > 0$, such that
$\{e_k, h_k\}$ is a basis of $H$.
Also we assume that
there is a constant $0<\theta<1$ such that
\begin{equation}
\label{c5}
B_\lambda :H_\theta \longrightarrow H \,\,\text{bounded, $\forall$
$\lambda\in \R^1$.}
\end{equation}
Under conditions (\ref{c4}) and (\ref{c5}), the operator
$-L_\lambda=A + B_\lambda$ is a sectorial operator.

Let $\{S_\lambda(t)\}_{t\ge 0}$ be an operator semi-group generated by
the equation (\ref{c1}).
Then the solution of (\ref{c1})  can be expressed
as $\psi(t, \psi_0) = S_\lambda(t)\psi_0,$  for any $ t\ge 0.$

\begin{definition}
\label{dfc1}
A set $\Sigma \subset H$ is called an invariant set of
(\ref{c1}) if $S(t) \Sigma = \Sigma$ for any $t\ge 0$. An
invariant set $\Sigma \subset H$ of (\ref{c1}) is called an
attractor if $\Sigma$ is compact, and there exists a neighborhood $W
\subset H$ of $\Sigma$ such that for any $\psi_0\in W$ we have
$$
\lim_{t\to \infty}\dist_H(\psi(t,\psi_0),\Sigma)= 0.
$$
\end{definition}

\begin{definition}
\label{dfc2}
\begin{enumerate}

\item We say that the solution of (\ref{c1})
bifurcates from $(\psi,\lambda) =
(0,\lambda_0)$ to an invariant set $\Omega_\lambda$, if there exists a
sequence of invariant sets $\{\Omega_{\lambda_n}\}$ of (\ref{c1})
such that $0 \notin \Omega_{\lambda_n}$,
$\lim_{n\to \infty} \lambda_n = \lambda_0$, and
$$ \lim_{n\to \infty} \max_{x\in \Omega_{\lambda_n}} |x| =0.$$

\item If the invariant sets $\Omega_\lambda$ are attractors of
(\ref{c1}), then the bifurcation is called attractor bifurcation.

\end{enumerate}
\end{definition}
\subsection{Main theorems}
In this article, we consider two cases:
\begin{align}
\label{c6} & \sigma >1 \qquad \text{and} && R_{c_1}\,\, \text{is
obtained only at} \,\, (j,k,l)=(j_1, 0, 1),
\,\, \\
\label{c7} &\sigma <1 \qquad \text{and} &&R_{c_2} \,\, \text{is
obtained only at} \,\, (j,k,l)=(j_2, 0, 1),
\end{align}
for some $j_1$, $j_2 \in \mathbb N$,
where $R_{c_1}$ and $R_{c_2}$ are defined in (\ref{e18})
 and (\ref{e22}) respectively.
In the above cases, $R_{c_1}$ and $R_{c_2}$ are given by the
following formulas:
\begin{align*}
& R_{c_1}=\frac{(j_1^2\alpha_1^2+\pi^2)^3}{j_1^2 \alpha_1^2}
  +\frac{\pi^2}{\sigma^2 Ro^2 j_1^2 \alpha_1^2},\\
& R_{c_2}=\frac{2(\sigma+1)(j_2^2\alpha_1^2+\pi^2)^3}{j_2^2\alpha_1^2}
   +\frac{2 \pi^2}{(\sigma+1)Ro^2 j_2^2 \alpha_1^2 }.
\end{align*}
\begin{remark}
{\rm
\begin{enumerate}
 \item Condition (\ref{c6}) guarantees that
     for $R\approx R_{c_1}$, the first
     eigenvalue of $L_{R}\mid_{H_1}$
     (resp., $L_{R}\mid_{\tilde{H}_1}$) is
     real and of multiplicity
     two (resp., one); see Remark~\ref{rme3}.
  \item Condition (\ref{c7}) guarantees that,
     for $R\approx R_{c_2}$, there exists
     only one simple pair of conjugate complex
     eigenvalues  of  $L_{R}\mid_{\tilde{H}_1}$
     crossing the imaginary axis; see Lemma~\ref{le6}.
  \item Condition (\ref{c6}) or (\ref{c7})
    can be satisfied easily; see Lemmas~\ref{le4} and ~\ref{le5}.
\end{enumerate}
}
\end{remark}

\begin{theorem}
\label{thc4}
Assume  (\ref{c6}). Then the following
assertions for Problem (\ref{b1})-(\ref{b4}) defined
in $H$ hold true.
\begin{enumerate}
\item If $R \le R_{c_1}$, the steady state $(U,T)=0$
      is locally asymptotically stable.

\item For $R > R_{c_1}$, the problem bifurcates from
      $((U,T),R)=(0,R_{c_1})$ to an attractor
      $\Sigma_{R}=S^1$, consisting of only steady state solutions.
\end{enumerate}
\end{theorem}

\begin{figure}
 \centering \includegraphics[height=.4\hsize]{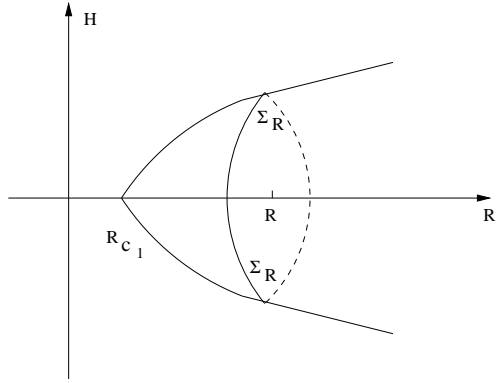}
 \caption{Bifurcation from $(0,R_{c_1})$ to an attractor
         $\Sigma_{R}$ for $R > R_{c_1}$. }
 \label{fig1}
\end{figure}

\begin{theorem}
\label{thc5}
Assume  (\ref{c7})
 and
 $$Ro^2 < \frac{(1-\sigma)\pi^2}{\sigma^2 (1+\sigma)(j_2^2\alpha_1^2 + \pi^2)^3}.$$
 The following statements are true.
 \begin{enumerate}
 \item For Problem (\ref{b1})-(\ref{b4}) defined in $H$, the steady
 state $(U,T)=0$ is locally asymptotically stable if $R<R_{c_2}$.
 \item For Problem (\ref{b1})-(\ref{b4}) defined in
$\tilde{H}$, a Hopf bifurcation occurs generically when $R$ crosses
$R_{c_2}$.
 \end{enumerate}
\end{theorem}

\section{Preliminaries}
\subsection{Attractor bifurcation theory}
Consider (\ref{c1}) satisfying (\ref{c2})  and (\ref{c3}).
We start with the Principle of Exchange of Stabilities (PES).
Let the eigenvalues (counting the multiplicity) of $L_\lambda$ be given by
$\beta_1(\lambda)$, $\beta_2(\lambda)$, $\cdots$.
Suppose that
\begin{equation}
\label{d1} Re\beta_i(\lambda)\begin{cases}
<0 \,\,\,\, \text{if}\,\,\,\, \lambda<\lambda_0,\\
=0 \,\,\,\, \text{if}\,\,\,\, \lambda=\lambda_0,\\
>0 \,\,\,\, \text{if}\,\,\,\, \lambda>\lambda_0,
\end{cases}
\qquad \text{if}\qquad 1\le i\le m,
\end{equation}

\begin{equation}
\label{d2} Re\beta_j(\lambda_0)<0,\qquad \text{if} \qquad m+1\le j.
\end{equation}

Let the eigenspace of $L_\lambda$ at $\lambda_0$ be
\[
E_0=\displaystyle{\bigcup_{1\le j \le m}\bigcup_{k=1}^{\infty}}\{u,v\in H_1\mid
(L_{\lambda_0}-\beta_j(\lambda_0))^k w=0, w=u+iv \}.
\]
It is known that $\dim E_0=m$.

\begin{theorem}[T. Ma and S. Wang \cite{b-book, mw-db1}]
\label{thd1}
Assume that the conditions (\ref{c2})-(\ref{c5}) and
(\ref{d1})-(\ref{d2}) hold true,
and $u=0$ is locally asymptotically
stable  for (\ref{c1})
at $\lambda=\lambda_0$. Then the following assertions hold true.

\begin{enumerate}
\item For $\lambda>\lambda_0$, (\ref{c1}) bifurcates
from $(u,\lambda)=(0,\lambda_0)$ to
attractors $\Sigma_\lambda$, having the same homology as  $S^{m-1}$,
 with $ m-1\le dim\Sigma_\lambda\le m$, which is connected
if $ m>1$;
\item For any $u_\lambda\in\Sigma_\lambda$, $u_\lambda$ can
be expressed as
\[
u_\lambda=v_\lambda+o(\|v_\lambda\|_{H_1}), \,\, v_\lambda\in E_0;
\]
\item There is an open set $U\subset H$ with $0\in U$ such that the
attractor $\Sigma_\lambda$ bifurcated from $(0,\lambda_0)$
attracts $U\backslash\Gamma$ in $ H$, where $\Gamma$ is the stable
manifold of $u=0$ with co-dimension m.
\end{enumerate}
\end{theorem}

\subsection{Center manifold theory}
A crucial ingredient for the proof of the main theorems using the above
attractor bifurcation theorem is an approximation
formula for center manifold functions; see  \cite{b-book}.

Let $H_1$  and $H$ be decomposed into
\begin{equation}
\label{d3}
H_1 = E^\lambda_1 \oplus E^\lambda_2,  \qquad  H = \widetilde E^\lambda_1 \oplus \widetilde E^\lambda_2,
\end{equation}
for $\la$ near $\lambda_0 \in \R^1$, where $E^\lambda_1$,
$E^\lambda_2$ are invariant subspaces of $L_\lambda$, such that
$\dim E^\lambda_1<\infty$, $\tilde E^\lambda_1 = E^\lambda_1$,
$\widetilde E^\lambda_2 =$ closure of $E^\lambda_2$ in $H$.
In addition,  $L_\lambda$ can be decomposed into $L_\lambda =
\cL^\lambda_1 \oplus \cL^\lambda_2$ such that for any $\lambda$
near $\lambda_0$,
\begin{equation}
\label{d4}
\begin{cases}
\cL^\lambda_1 = L_\lambda |_{E^\lambda_1} : E^\lambda_1
\longrightarrow \widetilde E^\lambda_1, & \\
\cL^\lambda_2 = L_\lambda|_{E^\lambda_2}:E^\lambda_2 \longrightarrow
\widetilde E^\lambda_2, &
\end{cases}
\end{equation}
where all eigenvalues of $\cL^\lambda_2$ possess negative real
parts, and the eigenvalues of $\cL^\lambda_1$ possess  nonnegative
real parts at $\lambda=\lambda_0$.
Furthermore, with $\mu < 1$ given by (\ref{c3}), let
$$E^\la_2(\mu)=\text{ closure of $E^\la_2$ in } H_\mu.
$$

By the classical center manifold theorem (see among others
\cite{henry,temam}),
 there exists a neighborhood of
$\lambda_0$ given by $|\lambda-\lambda_0|<\delta$ for some
$\delta>0$, a neighborhood $B_\lambda \subset E^\lambda_1$ of
$x=0$, and a $C^1$ center manifold
function $\Phi(\cdot,\lambda):B_\lambda \to
E^\lambda_2(\theta)$, called the center manifold function,
depending continuously on $\lambda$.
Then to investigate the
dynamic bifurcation of (\ref{c1}) it suffices to consider the
finite dimensional system as follows
\begin{equation}
\label{d5}
\frac{dx}{dt} = \cL^\lambda_1 x + g_1(x,\Phi_\lambda(x),\lambda),
\qquad x\in B_\lambda \subset E^\lambda_1.
\end{equation}
Hence, an approximation formula for the center manifold function
$\Phi_\lambda$ is crucial for the bifurcation and stability study.

Let the nonlinear operator $G$ be in the following form
\begin{equation}
\label{d6} G(u,\lambda)  = G_n(u,\lambda) + o(\|u\|^n),
\end{equation}
for some integer $n \ge 2$. Here $G_n:H_1 \times \cdots \times H_1
\longrightarrow H$ is a $n$-multilinear operator, and
$G_n(u,\lambda) = G_n(u,\cdots,u,\lambda).$

\begin{theorem}\cite{b-book}
\label{thd2}
Under the conditions (\ref{d3}), (\ref{d4}) and (\ref{d6}), the
center manifold function $\Phi(x,\lambda)$ can be expressed as
\begin{equation}
\label{d7} \Phi(x,\lambda) = (-\cL^\lambda_2)^{-1} P_2G_n(x,\lambda)
+ o(\|x\|^n) + O(|\text{\rm Re}\beta|\, \|x\|^n),
\end{equation}
where $\cL^\lambda_2$ is as in (\ref{d4}), $P_2:H\to
\widetilde E_2$ the canonical projection, $x\in E^\lambda_1$, and
$\beta= (\beta_1(\lambda),\cdots,\beta_m(\lambda))$ the
eigenvectors of $\cL^\lambda_1$.
\end{theorem}

\section{Eigenvalue Problem}
The eigenvalue problem of the linearized problem of (\ref{b1})-(\ref{b3}) is
given by
\be
\label{e1}
\left\{
\begin{aligned}
& \sigma( \Delta U-\nabla p)
  +\sigma R T e - \frac{1}{Ro}e \times U = \beta U,\\
& \Delta T +w = \beta T,\\
& \divv  U = 0,
\end{aligned}
\right.
\ee
supplemented with (\ref{b2}) and (\ref{b3}).
For $\psi=(U,T)$ satisfying (\ref{b2}) and (\ref{b3}), we expand the
field $\psi$ in Fourier series
\be
\label{e2}
\psi(x,y,z)=\sum_{j,k=-\infty}^{\infty}\psi_{jk}(z)e^{i(j\alpha_1 x
+k \alpha_2 y)}.
\ee
Plugging (\ref{e2}) into (\ref{e1}), we obtain the following
system of ordinary differential equations
\be
\label{e3}
\left\{
\begin{aligned}
& \sigma(D_{jk} u_{jk} - ij\alpha_1 p_{jk})+\frac{1}{Ro}v_{jk}=\beta u_{jk},\\
& \sigma(D_{jk} v_{jk} - ik\alpha_2 p_{jk})-\frac{1}{Ro}u_{jk}=\beta v_{jk},\\
& D_{jk} w_{jk}-p_{jk}'+R T_{jk} =
   \sigma^{-1}\beta w_{jk},\\
& D_{jk} T_{jk} + w_{jk} = \beta T_{jk},\\
& ij\alpha_1 u_{jk}+ik\alpha_2 v_{jk} + w_{jk}'=0,\\
& u_{jk}' \mid_{z=0,1} = v_{jk}' \mid_{z=0,1}= w_{jk}\mid_{z=0,1}=
T_{jk}\mid_{z=0,1}=0,
\end{aligned}
\right.
\ee
for $j, k \in \mathbb Z$, where $'=d/dz$, $D_{jk}=d^2/dz^2-\alpha_{jk}^2$
and $\alpha_{jk}^2=j^2\alpha_1^2+k^2\alpha_2^2$. If $w_{jk} \ne 0$,
(\ref{e3}) can be reduced to  a single equation for $w_{jk}(z)$:
\begin{align}
\label{e4} & \{ (D_{jk}-\beta)(\sigma D_{jk}-\beta)^2 D_{jk} \\
     & \qquad + \frac{1}{Ro^2}(D_{jk}-\beta)(D_{jk}+\alpha_{jk}^2)
      + \sigma R \alpha_{jk}^2(\sigma D_{jk}-\beta) \} w_{jk} =0,
    \nonumber \\
\label{e5}
& w_{jk}=w_{jk}^{''}=w_{jk}^{(4)}=w_{jk}^{(6)}=0 \qquad \text{at} \qquad z=0,1,     \end{align}
for $j,k \in \mathbb Z$. Thanks to (\ref{e5}), $w_{jk}$
can be expanded in a  Fourier sine series
\be
\label{e6}
w_{jk}(z)=\sum_{l=1}^{\infty}w_{jkl}\sin l\pi z,
\ee
for $(j,k)\in \mathbb Z \times \mathbb Z$.
Substituting (\ref{e6}) into (\ref{e4}),
we see that the
 eigenvalues $\beta$ of the problem (\ref{e1}) satisfy
the cubic equations
\begin{align}
\label{e7}& \beta^3  + (2\sigma +1) \gamma_{jkl}^2 \beta^2 +
[(\sigma^2+2\sigma)\gamma_{jkl}^4 + \frac{l^2 \pi^2}{Ro^2 \gamma_{jkl}^2}
 -\sigma R \frac{\alpha_{jk}^2}{\gamma_{jkl}^2}]\beta \\
&\,\,\,  +  \sigma^2 \gamma_{jkl}^6 -\sigma^2 R \alpha_{jk}^2
+\frac{l^2 \pi^2}{Ro^2}=0, \nonumber
\end{align}
for $j,k\in \mathbb Z$ and $l\in \mathbb N$,
where $\gamma_{jkl}^2=\alpha_{jk}^2 + l^2\pi^2$.
In the following discussions, we let
\begin{equation}
\label{e8}
\begin{aligned}
& g_{jkl}(\beta)=(\beta+\gamma_{jkl}^2)[(\beta + \sigma \gamma_{jkl}^2)^2
         + l^2 \pi^2 Ro^{-2} \gamma_{jkl}^{-2}],\\
& h_{jkl}(\beta)= \sigma R \alpha_{jk}^2 \gamma_{jkl}^{-2}
 (\beta + \sigma \gamma_{jkl}^2), \\
& f_{jkl}(\beta)=g_{jkl}(\beta) - h_{jkl}(\beta), \\
\end{aligned}
\end{equation}
and $\beta_{jkl1}(R)$, $\beta_{jkl2}(R)$ and $\beta_{jkl3}(R)$
be the zeros of
$f_{jkl}$ with
$$
 Re(\beta_{jkl1}) \ge Re(\beta_{jkl2}) \ge Re(\beta_{jkl3}).
$$
\subsection{Eigenvectors}
In the following discussions,
we consider the following index sets:
\begin{align*}
 & \Lambda_1= \{(j,k,l) \in \mathbb Z^2 \times \mathbb N  \mid
      j \ge 0, (j,k) \ne (0,0)\},\\
 & \Lambda_2= \{(j,k,l) \in \mathbb Z^2 \times \{0\} \mid
         j \ge 0, (j,k) \ne (0,0)\},\\
 & \Lambda_3= \{ (j,k,l) \in \{(0,0)\} \times \mathbb N\},\\
 & \Lambda = \Lambda_1 \cup \Lambda_2 \cup \Lambda_3.
\end{align*}

\medskip

1. For $(j,k,0) \in \Lambda_2$, we define
\begin{align*}
 & \psi^{\beta_{jk0}}_1= ( k \alpha_2 \sin
  (j \alpha_1 x +k\alpha_2 y), -j\alpha_1 \sin(j \alpha_1 x + k \alpha_2 y)
   ,0,0)^t,\\
& \psi^{\beta_{jk0}}_2= (-k \alpha_2 \cos (j\alpha_1 x + k \alpha_2 y
   ), j \alpha_1 \cos (j \alpha_1 x + k\alpha_2 y),0 ,0 )^t,\\
& E_{jk0}=span \{\psi^{\beta_{jk0}}_1, \psi^{\beta_{jk0}}_2\},\\
& \beta_{\Lambda_2}= \cup_{(j,k,0)\in \Lambda_2} \{\beta_{jk0}\},
 \end{align*}
where $\beta_{jk0}=- \sigma \gamma_{jk0}^2
=-\sigma \alpha_{jk}^2= -\sigma(j^2 \alpha_1^2 + k^2 \alpha_2^2).$ It
is not hard to see that
$L_R (\psi^{\beta_{jk0}}_1 )=\beta_{jk0}\psi^{\beta_{jk0}}_1$ and
$L_R (\psi^{\beta_{jk0}}_2 )=\beta_{jk0}\psi^{\beta_{jk0}}_2$.

\medskip

2. For $(0,0,l) \in \Lambda_3$, we define
\begin{align*}
& \psi^{\beta_{00l1}}=(0,0,0,\sin l\pi z)^t,
&& \psi^{\beta_{0012}}=(\cos l\pi z,0,0,0)^t, \\
& \psi^{\beta_{00l3}}=(0,\cos l\pi z,0,0)^t,
&& E_{00l}= span \{ \psi^{\beta_{00l1}},
   \psi^{\beta_{00l2}}, \psi^{\beta_{00l3}}\}, \\
& \beta_{\Lambda_3}=\cup_{l=1}^{\infty} \cup_{q=1}^{3} \{\beta_{00lq}\},
&& \beta_{\tilde{\Lambda}_3}=\cup_{l=1}^{\infty} \{\beta_{00l1} \},
\end{align*}
where $\beta_{00l1}= - \gamma_{00l}^2 = - l^2 \pi^2$,
$\beta_{0012}= - \sigma \gamma_{00l}^2 - \frac{1}{Ro} i$ and
$\beta_{00l3}= - \sigma \gamma_{00l}^2 + \frac{1}{Ro} i$.
It is easy to check that
\begin{align*}
& L_R (\psi^{\beta_{00l1}})=\beta_{0011}\psi^{\beta_{00l1}},\\
& L_R (\psi^{\beta_{00l2}})=
  - \sigma \gamma_{00l}^2 \psi^{\beta_{00l2}}
  -\frac{1}{Ro} \psi^{\beta_{00l3}},\\
& L_R (\psi^{\beta_{00l3}})=
    \frac{1}{Ro} \psi^{\beta_{00l2}}
  - \sigma \gamma_{00l}^2 \psi^{\beta_{00l3}}.
   \end{align*}

\medskip
3. For $(j,k,l) \in \Lambda_1$, we define
\begin{align*}
& \phi_{jkl}^1=( -\frac{j\alpha_1 l\pi}{\alpha_{jk}^2}
               \sin(j\alpha_1 x+ k\alpha_2 y )\cos l\pi z,
                -\frac{k\alpha_2 l\pi}{\alpha_{jk}^2}
               \sin(j\alpha_1 x+ k\alpha_2 y )\cos l\pi z,\\
           & \qquad \qquad \qquad \qquad
            \cos(j\alpha_1 x+k \alpha_2 y )\sin l\pi z,
            0)^t , \\
& \phi_{jkl}^2=(\frac{k\alpha_2 l\pi}{\alpha_{jk}^2}
               \sin(j\alpha_1 x+ k\alpha_2 y )\cos l\pi z,
                -\frac{j\alpha_1 l\pi}{\alpha_{jk}^2}
               \sin(j\alpha_1 x+ k\alpha_2 y )\cos l\pi z,0,0),\\
& \phi_{jkl}^3=(0,0,0,\cos (j\alpha_1 x + k \alpha_2 y)
                \sin l\pi z)^t,\\
& \phi_{jkl}^4=( \frac{j\alpha_1 l\pi}{\alpha_{jk}^2}
            \cos (j\alpha_1 x+ k\alpha_2 y) \cos l\pi z,
            \frac{k\alpha_2 l\pi}{\alpha_{jk}^2}
            \cos (j\alpha_1 x+ k\alpha_2 y) \cos l\pi z,\\
& \qquad \qquad \qquad \qquad
            \sin (j\alpha_1 x+k\alpha_2 y) \sin l\pi z,
            0 )^t ,\\
& \phi_{jkl}^5= (-\frac{k\alpha_2 l\pi}{\alpha_{jk}^2}
            \cos (j\alpha_1 x+ k\alpha_2 y) \cos l\pi z,
             \frac{j\alpha_1 l\pi}{\alpha_{jk}^2}
            \cos (j\alpha_1 x+ k\alpha_2 y) \cos l\pi z,
            0,0)^t, \\
& \phi_{jkl}^6 = (0,0,0, \sin(j \alpha_1 x + k \alpha_2 y)
\sin l \pi z)^t,\\
& E_{jkl}^1 = span \{\phi_{jkl}^1, \phi_{jkl}^2, \phi_{jkl}^3 \},
\qquad  E_{jkl}^2 = span \{\phi_{jkl}^4, \phi_{jkl}^5, \phi_{jkl}^6 \},\\
& E_{jkl}=E_{jkl}^1 \oplus E_{jkl}^2,
 \qquad  \beta_{\Lambda_1}=
  \cup_{(j,k,l) \in \Lambda_1} \cup_{q=1}^3 \{ \beta_{jklq}\}.
\end{align*}
It is easy to check that $E_{jkl}^1$ and $E_{jkl}^2$  are invariant
subspaces of the linear operator $L_R$ respectively, i.e., $ L_R (E_{jkl}^1) \subset  E_{jkl}^1$ and
$L_R (E_{jkl}^2) \subset E_{jkl}^2$. The characteristic polynomial
of $L_R \mid_{E_{jkl}^1}$ (resp., $L_R \mid_{E_{jkl}^2}$)
 is given by $f_{jkl}$ as
defined in (\ref{e8}).
Since $E_{jkl}^1$ (resp.,$E_{jkl}^2$) is of dimension three, the (generalized)
eigenvectors of $L_R\mid_{E_{jkl}^1}$,
$\cup_{q=1}^{3}\{\psi^{\beta_{jklq}}_1\}$
($\cup_{q=1}^{3}\{\psi^{\beta_{jklq}}_2\}$),
form a basis of $E_{jkl}^1$ (resp., $E_{jkl}^2$), i.e.,
span$\{\cup_{q=1}^{3}\{\psi^{\beta_{jklq}}_1\}\}=E_{jkl}^1$
(resp., span$\{\cup_{q=1}^{3}\{\psi^{\beta_{jklq}}_2\}\}=E_{jkl}^2$).
 If $\beta_{jklq}$ is a real zero of $f_{jkl}$,
the eigenvector corresponding to $\beta_{jklq}$ in $E_{jkl}^1$ \
(resp., $E_{jkl}^2$) is
given by

\begin{align}
\label{e9}
& \psi^{\beta_{jklq}}_1=\phi^1_{jkl}+A_1(\beta_{jklq})\phi^2_{jkl}
                +A_2(\beta_{jklq})\phi^3_{jkl},
           \\
& ( \psi^{\beta_{jklq}}_2=\phi^4_{jkl}+A_1(\beta_{jklq})\phi^5_{jkl}
                +A_2(\beta_{jklq})\phi^6_{jkl} ),
\nonumber
\end{align}

where
\be
\label{e10}
A_1(\beta)=\frac{-1}{Ro(\beta + \sigma \gamma_{jkl}^2)},
\qquad A_2(\beta)=\frac{1}{\beta+\gamma_{jkl}^2}.
\ee
If $\beta_{jklq_1}=\bar{\beta}_{jklq_2}$ (imaginary
numbers) are zeros of $f_{jkl}$, the (generalized)
 eigenvectors  corresponding to $\beta_{jklq_1}$
and $\beta_{jklq_2}$ in $E_{jkl}^1$
(resp., $E_{jkl}^2$) are
given by
\begin{equation}
\label{e11}
\begin{aligned}
& \psi^{\beta_{jklq_1}}_1=\phi_{jkl}^1
   +R_1(\beta_{jklq_1})\phi_{jkl}^2
   +R_2(\beta_{jklq_1})\phi_{jkl}^3, \\
&  \psi^{\beta_{jklq_2}}_1=I_1(\beta_{jklq_1})\phi_{jkl}^2
   +I_2(\beta_{jklq_1})\phi_{jkl}^3,
\end{aligned}
\end{equation}
\begin{eqnarray*}
\left(
\begin{aligned}
& \psi^{\beta_{jklq_1}}_2=\phi_{jkl}^4
   +R_1(\beta_{jklq_1})\phi_{jkl}^5
   +R_2(\beta_{jklq_1})\phi_{jkl}^6, \\
&  \psi^{\beta_{jklq_2}}_2=I_1(\beta_{jklq_1})\phi_{jkl}^5
   +I_2(\beta_{jklq_1})\phi_{jkl}^6,
\end{aligned}
\right),
\end{eqnarray*}
where
\begin{equation}
\label{e12}
\begin{aligned}
& R_1 (\beta) = Re(A_1(\beta)), \qquad R_2 (\beta)= Re(A_2(\beta)),\\
& I_1 (\beta) = Im(A_1(\beta)), \qquad I_2 (\beta)= Im(A_2(\beta)).
\end{aligned}
\end{equation}

The dual vector corresponding to $\psi^{\beta_{jklq}}_1$
(resp., $\psi^{\beta_{jklq}}_2$) is given by

\begin{align}
\label{e13}
& \Psi^{\beta_{jklq}}_1=\phi^1_{jkl}+C_1(\beta_{jklq})\phi^2_{jkl}
                +C_2(\beta_{jklq})\phi^3_{jkl},
           \\
& (\Psi^{\beta_{jklq}}_2=\phi^4_{jkl}+C_1(\beta_{jklq})\phi^5_{jkl}
                +C_2(\beta_{jklq})\phi^6_{jkl}),
\nonumber
\end{align}
where
\be
\label{e14}
C_1(\beta)=\frac{1}{Ro(\beta + \sigma \gamma_{jkl}^2)},
\qquad C_2(\beta)=\frac{\sigma R}{\beta+\gamma_{jkl}^2}.
\ee
The dual vector $\Psi^{\beta_{jklq}}_1$ (resp., $\Psi^{\beta_{jklq}}_2$)
satisfies
\begin{align}
\label{e15}
& <\psi^{\beta_{jklq^{*}}}_1 ,\Psi^{\beta_{jklq}}_1 >_H =0 \qquad
 ( <\psi^{\beta_{jklq^{*}}}_2 ,\Psi^{\beta_{jklq}}_2 >_H = 0),
\end{align}
for $q^* \ne q$.

\medskip

We note that $E_{j_1 k_1 l_1}$ is orthogonal to
$E_{j_2 k_2 l_2}$  for $(j_1, k_1, l_1) \ne (j_2, k_2, l_2)$
and $E_{jkl}^1$ is orthogonal to $E_{jkl}^2$ for
$(j,k,l) \in \Lambda_1$. Hence the dual vector
$\Psi^{\beta_{jklq}}_1$ (resp., $\Psi^{\beta_{jklq}}_2$)
satisfies
\begin{align}
\label{e16}
& <\psi, \Psi^{\beta_{jklq}}_1>_H = 0 \qquad
 \text{for} \qquad  \psi \in (\cup_{(j^*, k^*, l^*)\ne(j,k,l)}
E_{j^* k^* l^*})\cup E_{jkl}^2\\
& ( <\psi, \Psi^{\beta_{jklq}}_2>_H = 0 \qquad
 \text{for} \qquad \psi \in (\cup_{(j^*, k^*, l^*)\ne(j,k,l)}
E_{j^* k^* l^*})\cup E_{jkl}^1).
\nonumber
\end{align}

In view of the Fourier expansion, we see that
$\cup_{(j,k,l)\in \Lambda} E_{jkl}$ is a basis of $H_1$
and $(\cup_{(j,k,l) \in \Lambda_1} E_{jkl}^1)\cup
(\cup_{(j,k,0)\in \Lambda_2} \{\psi^{\beta_{jk0}}_1 \})
 \cup (\cup_{(0,0,l)\in \Lambda_3} \{\psi^{\beta_{00l1}} \}) $
 is a basis of $\tilde{H}_1$. Hence, by the discussion above, we
have the following conclusions.

\medskip

a) The set
  $\beta_{H_1}=\beta_{\Lambda_1}
  \cup \beta_{\Lambda_2} \cup \beta_{\Lambda_3}$
    consists of all eigenvalues of $L_R \mid_{H_1}$,
   and the (generalized) eigenvectors of $L_R \mid_{H_1}$
 form a basis of $H_1$.

\medskip

b) The set $\beta_{\tilde{H}_1}=\beta_{\Lambda_1} \cup \beta_{\Lambda_2}
    \cup \beta_{\tilde{\Lambda}_3} $
   consists of all eigenvalues of  $L_R \mid_{\tilde{H}_1}$, and
  the (generalized) eigenvectors of  $L_R \mid_{\tilde{H}_1}$
  form a basis of $\tilde{H}_1$.

\medskip

c) $Re(\beta) < 0$ for each
   $\beta \in  \beta_{\Lambda_2} \cup \beta_{\Lambda_3}$.

\begin{lemma}
\label{le1}
 If $R$ is small, then $ Re(\beta_{jklq}(R)) < 0 $
for each $\beta_{jklq} \in \beta_{\Lambda_1}$.
\end{lemma}
\begin{proof}
Plugging $\beta = \gamma_{jkl}^2 \beta^*$ into $f_{jkl}$, we get
$ f_{jkl}(\beta) =  \gamma_{jkl}^6 \tilde{f}_{jkl}(\beta^*)$,
where
 $$
\tilde{f}_{jkl}(\beta^*)= (\beta^*+1)(\beta^*+\sigma)^2
 + \frac{l^2 \pi^2}{ \gamma_{jkl}^6 Ro^2}(\beta^*+1)-
   \sigma R \frac{\alpha_{jk}^2}{\gamma_{jkl}^6}(\beta^*+\sigma).
$$
Hence, we only need to show that the real part of each
zero of $\tilde{f}_{jkl}$
is strictly negative when $R$ is small. We observe that
 $\tilde{f}_{jkl}(\beta^*)>0$ for all $\beta^* \ge  0$  provided
 $ R < 1 + \sigma^{-1}$. Therefore, if all zeros of $\tilde{f}_{jkl}$ are
real numbers, we are done.

 For the case where only one of the zeros of
 $\tilde{f}_{jkl}$ is real, this real zero, $\beta^*_1$, is a perturbation
of $-1$. There exists an $\epsilon$ ( depending on $\sigma$ only)
such that $ -(1+2 \sigma) < \beta^*_1 < 0 $ provided $ R < \epsilon$. This
makes the real part of the other two zeros of $\tilde{f}_{jkl}$
strictly negative and the proof is complete.
\end{proof}

\subsection{Characterization of Critical Rayleigh Numbers}
Based on the above discussion, we know that
only the eigenvalues in $\beta_{\Lambda_1}$ depend on the Rayleigh
number $R$. Hence, to study the Principle of Exchange
of Stabilities for problem (\ref{e1}),
it suffices to focus the problem on the set $\beta_{\Lambda_1}$.
We proceed with the following two cases.

{\sc Case 1.}  $\beta=0$ is a zero
      of $f_{jkl}$ if and only if the constant term
      of the polynomial $f_{jkl}$ is $0$. In this
      case, we have
      \begin{equation}
      \label{e17}
      R = \frac{\gamma_{jkl}^6}{\alpha_{jk}^2}
        + \frac{l^2 \pi^2}{\sigma^2 Ro^2 \alpha_{jk}^2}
        \ge \frac{(\alpha_{jk}^2+ \pi^2)^3}{\alpha_{jk}^2}+
        \frac{ \pi^2}{\sigma^2 Ro^2 \alpha_{jk}^2}.
      \end{equation}
     Hence the critical Rayleigh number $R_{c_1}$ is given by
      \be
    \label{e18}
    R_{c_1}= \min_{(j,k,l)\in \Lambda_1}
           \{ \frac{\gamma_{jkl}^6}{\alpha_{jk}^2}
            + \frac{l^2 \pi^2}{\sigma^2 Ro^2 \alpha_{jk}^2} \}
           = \frac{\gamma_{j_1 k_1 1}^6}{\alpha_{j_1 k_1}^2}
            + \frac{ \pi^2}{\sigma^2 Ro^2 \alpha_{j_1 k_1}^2},
    \ee
    for some $(j_1, k_1, 1 )\in \Lambda_1$.

{\sc Case 2.}
A careful analysis on (\ref{e7}) shows that $\beta=ai$
      ($a \ne 0$), a purely imaginary number, is a zero
      of $f_{jkl}$  if and only if the following two equations hold true:
     \begin{align*}
     & (\sigma^2 + 2\sigma)\gamma_{jkl}^4 +
       \frac{l^2 \pi^2}{Ro^2 \gamma_{jkl}^2}
       -\sigma R \frac{\alpha_{jk}^2}{\gamma_{jkl}^2}
       >0, \\
     & (2\sigma +1 )\gamma_{jkl}^2 [
        (\sigma^2 + 2\sigma)\gamma_{jkl}^4
         + \frac{l^2 \pi^2}{ Ro^2 \gamma_{jkl}^2}
           -\sigma R \frac{\alpha_{jk}^2}{\gamma_{jkl}^2}] \\
& \quad         = \sigma^2 \gamma_{jkl}^6 -\sigma^2 R \alpha_{jk}^2
           +\frac{l^2 \pi^2}{Ro^2}.
     \end{align*}
In this case,  we have
     \begin{align}
&     \label{e19}
      R = \frac{2(\sigma+1)\gamma_{jkl}^6}{\alpha_{jk}^2}+
           \frac{2 l^2 \pi^2}{(\sigma +1) Ro^2 \alpha_{jk}^2}, \\
&     \label{e20}
        R < \frac{(\sigma+2)\gamma_{jkl}^6}{\alpha_{jk}^2}
             + \frac{l^2 \pi^2}{\sigma Ro^2 \alpha_{jk}^2}.
     \end{align}
     Plugging (\ref{e20}) into (\ref{e19}), we derive an upper
     bound  for $Ro^2$,
    \be
    \label{e21}
     Ro^2 < \frac{(1-\sigma)l^2 \pi^2}{ \sigma^2(1+\sigma) \gamma_{jkl}^6  },
    \ee
    which could only hold true  when $\sigma < 1$.

\begin{figure}
 \centering \includegraphics[height=.5\hsize]{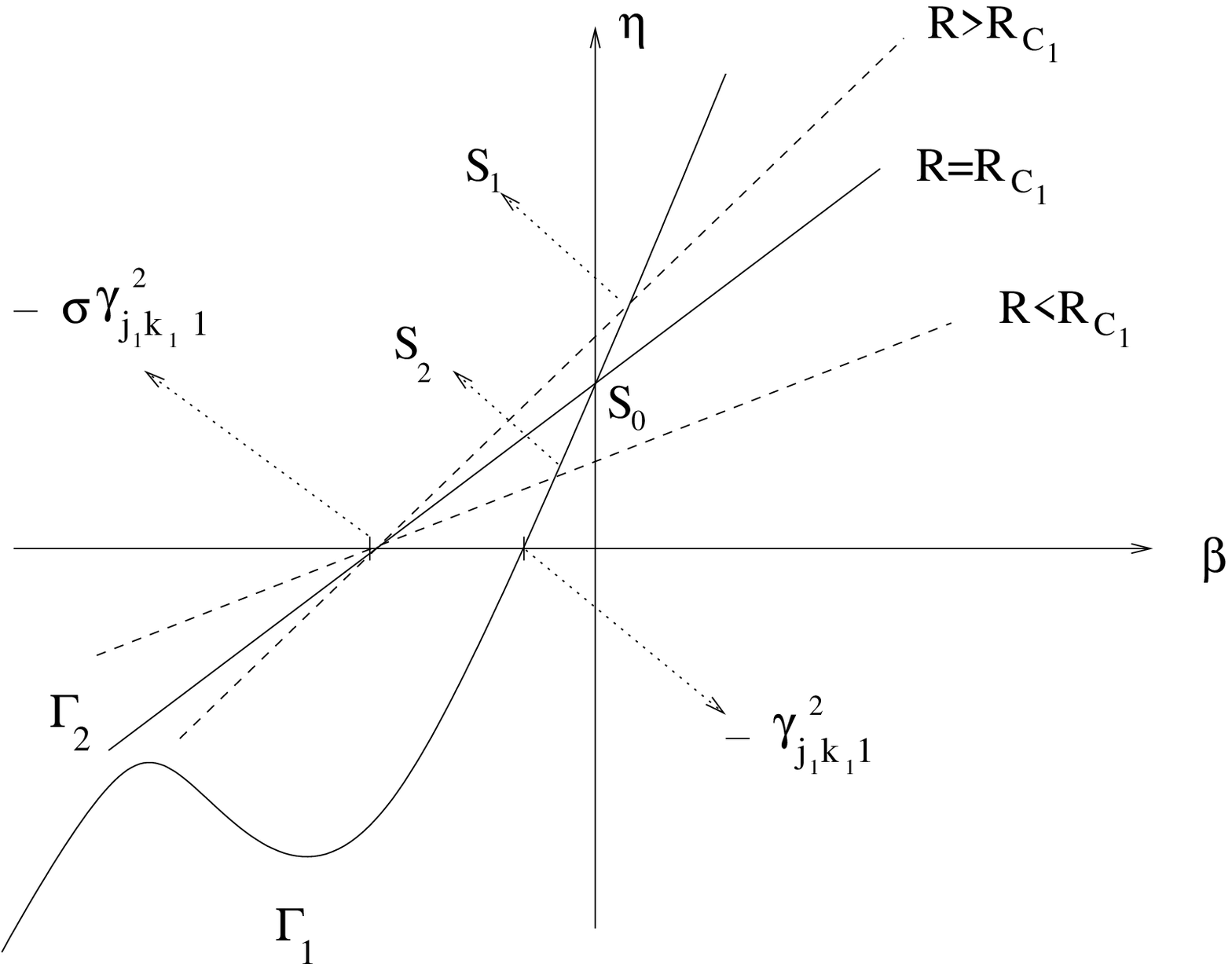}
 \caption{}
\label{fig2}
\end{figure}

As in  Case 1, the minimum of the right hand side
    of (\ref{e19}) is always obtain at $l=1$. Hence the critical
    Rayleigh number $R_{c_2}$  is given by
    \begin{align}
    \label{e22}
    R_{c_2}=& \min_{(j,k,l)\in \Lambda_1}
           \{\frac{2(\sigma+1)\gamma_{jkl}^6}{\alpha_{jk}^2}+
             \frac{2 l^2 \pi^2}{(\sigma +1) Ro^2 \alpha_{jk}^2}
              \}\\
           =& \frac{2(\sigma+1)\gamma_{j_2 k_2 1}^6}{\alpha_{j_2 k_2}^2}+
             \frac{2  \pi^2}{(\sigma +1) Ro^2 \alpha_{j_2 k_2}^2},
      \nonumber
    \end{align}
    for some $(j_2, k_2, 1)\in \Lambda_1$.
    In the case of $\sigma < 1$, (\ref{e21}) with $l=1$ implies
    $R_{c_2}$ is smaller than $R_{c_1}$.
     Hence, for Problem (\ref{b1})-(\ref{b4}), $R_{c_1}$ is the first critical
    Rayleigh number  if $\sigma >1$ and
    $R_{c_2}$ is the first critical
    Rayleigh number if $\sigma < 1$.
    Therefore, the Principle of Exchange of Stabilities
    is given by Lemma~\ref{le2} and Lemma~\ref{le6}.
    \begin{lemma}
    \label{le2}
    For fixed $\sigma > 1$ and $Ro > 0$,
    suppose that $(\alpha_{jk}^2,l)=(\alpha_{j_1 k_1}^2,1)$
    minimizes the right
    hand side of (\ref{e17}), then
    \be
    \label{e23}
    \beta_{j_1 k_1 11}(R) \begin{cases}
<0 \,\,\,\, \text{if}\,\,\,\, R < R_{c_1}\\
=0 \,\,\,\, \text{if}\,\,\,\, R = R_{c_1}\\
>0 \,\,\,\, \text{if}\,\,\,\, R > R_{c_1}
                   \end{cases},
    \ee
    \be
    \label{e24}
    Re \beta_{jklq}(R) < 0 \qquad \text{for} \qquad
   (\alpha_{jk}^2, l) \ne (\alpha_{j_1k_1}^2, 1), \,q=1,2,3, \,\,
   R \,\, \text{near} \,\, R_{c_1}.
    \ee
     \end{lemma}
\begin{proof}
By the above discussion, we only need to show that the first eigenvalue
crosses the imaginary axis. We note that $f_{j_1 k_1 1}(\beta)=0$
is equivalent to $g_{j_1 k_1 1}(\beta)=h_{j_1 k_1 1}(\beta)$, i.e.,
\begin{equation}
\label{e25}
 (\beta+\gamma_{j_1 k_1 1}^2)[(\beta + \sigma \gamma_{j_1 k_1 1}^2)^2
         + l^2 \pi^2 Ro^{-2} \gamma_{j_1 k_1 1}^{-2}]=
\sigma R \alpha_{j_1 k_1}^2 \gamma_{j_1 k_1 1}^{-2}
 (\beta + \sigma \gamma_{j_1 k_1 1}^2).
\end{equation}
 We see that both $g_{j_1 k_1 1}$ and
 $h_{j_1 k_1 1}$ are strictly increasing
for $\beta > -\gamma_{j_1 k_1 1}^2$ ( since $\sigma > 1$ ).
Let $\Gamma_1$ be the graph of $\eta=g_{j_1 k_1 1}(\beta)$
and $\Gamma_2$  be the graph of $\eta=h_{j_1 k_1 1}(\beta)$
as shown in Figure~\ref{fig2}. When $R=R_{c_1}$, Point $S_0$, the intersecting
point of $\Gamma_1$ and $\Gamma_2$ corresponding to
$\beta_{j_1 k_1 1}(R)$ ( i.e., the $\beta$
coordinate of $S_0$ is $\beta_{j_1 k_1 1}(R)$), is on the $\eta$ axis.
When $R$ increases (resp., decreases), $S_0$ becomes $S_1$ ( resp., $S_2$).
This proves (\ref{e23}) and the proof is complete.
\end{proof}

\begin{remark}
\label{rme3}
{ \rm
\begin{enumerate}
\item In the proof of Lemma~\ref{le2}, as shown by (\ref{e25})
and Figure~\ref{fig2},
we see that, for $R \approx R_{c_1}$, the first eigenvalue
 $\beta_{j_1 k_1 1 1}$ is a simple zero of $f_{j_1 k_1 1}(\beta)$.
We have seen in Section 5.1 that there are eigenvectors
$\psi^{\beta_{j_1 k_111}}_1 \in E_{j_1 k_1 l}^1$  and
$\psi^{\beta_{j_1 k_111}}_2 \in E_{j_1 k_1 l}^2$ corresponding to
$\beta_{j_1 k_111}$. Therefore, the multiplicity of the first
eigenvalue of $L \mid_{H_1}$ (resp., $L \mid_{\tilde{H}_1}$) is
$m_{H_1}=2m$ (resp.,  $m_{\tilde{H}_1}=m$ ),
 where $m$ is the number of
 $(j,k,1)$'s  ($ \in \Lambda_1$) satisfying
$\alpha_{jk}^2=\alpha_{j_1 k_1}^2$.
Hence,  Condition (\ref{c6}) guarantees that,
     for $R\approx R_{c_1}$, the first
     eigenvalue of $L_{R}\mid_{H_1}$
     (resp., $L_{R}\mid_{\tilde{H}_1}$) is
     real and of multiplicity
     two (resp., one).
\item For the classical B\'enard problem without rotation,
      the second term on the right
      hand side of (\ref{e17}), hence the second term on
       the right hand side of
      (\ref{e18}), is not presented.
        Therefore, the first critical
      Rayleigh number of the classical B\'enard
      problem depends
      only on the aspect of ratio; while
      the first critical
      Rayleigh number of the rotating
      problem depends on the aspect
      of ratio, the Prandtl number and the
      Rossby number. And it is clear
      that the first critical Rayleigh
      number of fast rotating flows
      is remarkably larger than
     the first critical Rayleigh number of
     the classical B\'enard problem. This
     indicates that the rotating flows
      are much more stable than the
      non-rotating flows.
\item  $R_{c_1}$ is the first Critical
Rayleigh number if the Prandtl number is greater than one. For the
case where the Prandtl number is smaller than one, $R_{c_2}$ is the
first Critical Rayleigh number and, in general, there are a few
critical values between $R_{c_2}$ and $R_{c_1}$ .
       \end{enumerate}
 }
\end{remark}

\begin{figure}
 \centering \includegraphics[height=.5\hsize]{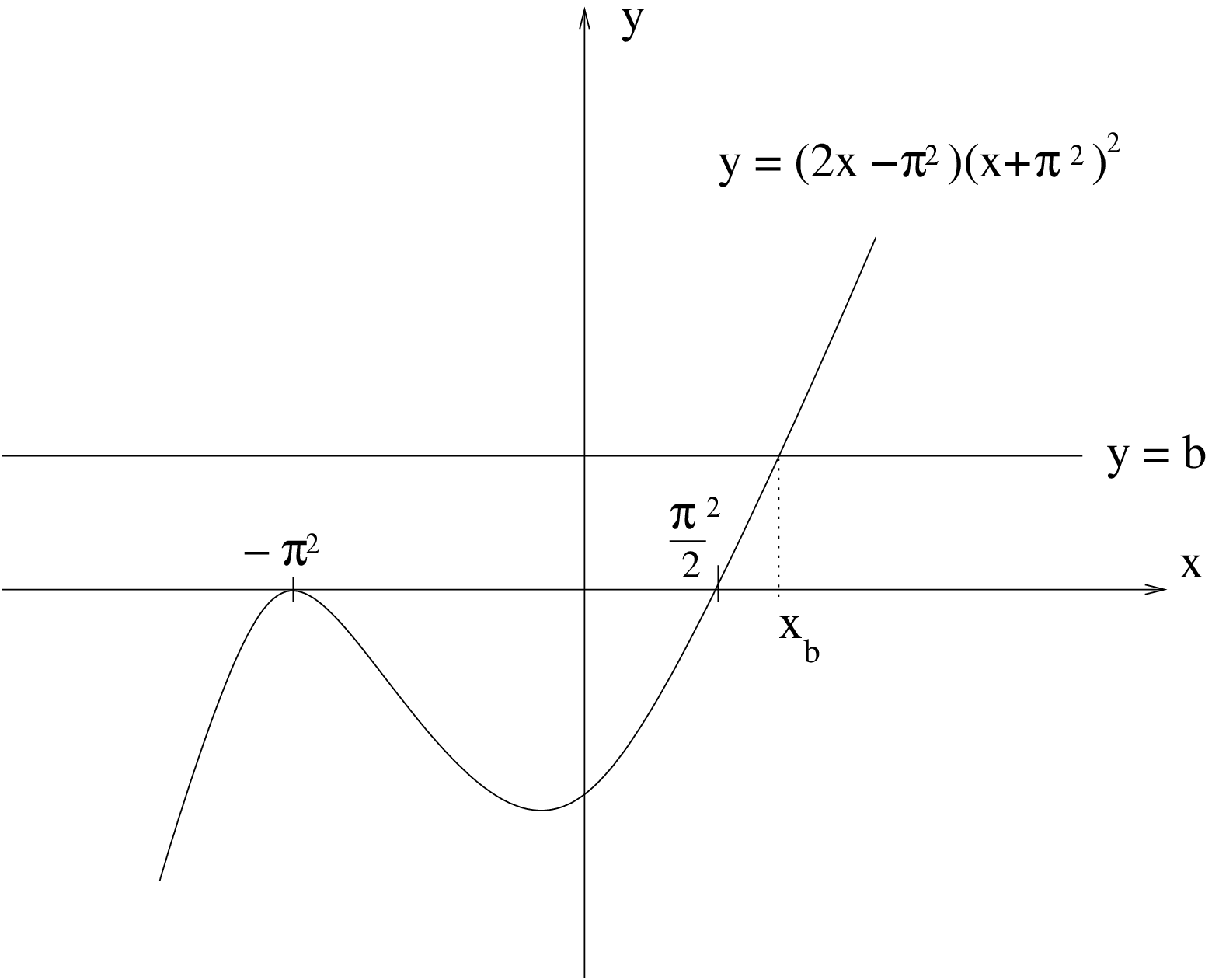}
 \caption{}
\label{fig3}
\end{figure}

For $x>0$, $b \ge 0$, we define
\begin{equation}
\label{e26}
 f_{b}(x)=\frac{(x+\pi^2)^3+b}{x}.
\end{equation}
Let $x=\alpha_{jk}^2$, then the right hand side of (\ref{e18})
could be expressed as $f_{b_1}(x)$, where
$b_1=\frac{\pi^2}{\sigma^2 Ro^2}$; and the second line of
(\ref{e22}) could be expressed as $2(\sigma+1)f_{b_2}(x)$, where
$b_2=\frac{\pi^2}{(\sigma+1)^2 Ro^2}$.
 Consider
\begin{equation}
\label{e27}
f^{'}_{b}(x)= \frac{(2x-\pi^2)(x+\pi^2)^2-b}{x^2}.
\end{equation}
 As shown in Figure~\ref{fig3},
it is easy to see that
\begin{enumerate}
\item[a)]   for $x \in (0,\infty)$, $f_{b}(x)$ has
only one critical number $x_{b}$,
\item[b)]   $f^{'}_{b}(x)<0$ if $x < x_{b}$,
\item[c)]   $f^{'}_{b}(x)>0$ if $x > x_{b}$,
\item[d)]    $f_{b}(x_{b})$ is the global
           minimum of $f_{b}(x)$, and
\item[e)]  $x_{b}$ is strictly increasing in $b$
           , hence, $x_{b_1} > x_{b_2} > \frac{\pi^2}{2}$.
\end{enumerate}

In Lemmas 5.4 and 5.5, we consider the following different conditions
\begin{align}
\label{e28}
& x_{b_1} \le \alpha_1^2 < \alpha_2^2,\\
\label{e29}
& \alpha_1^2 \le \frac{1}{5} x_{b_1} < 2 x_{b_1}< \alpha_{2}^2,\\
\label{e30}
& x_{b_2} \le \alpha_1^2 < \alpha_2^2,\\
\label{e31}
& \alpha_1^2 \le \frac{1}{5} x_{b_2} < 2 x_{b_2}< \alpha_{2}^2.
\end{align}

\begin{lemma}
\label{le4}
\begin{enumerate}
\item Condition (\ref{c6}) holds true under the assumption (\ref{e28}).
\item Generically, Condition (\ref{c6}) holds true under the assumption
      (\ref{e29}).
\end{enumerate}
\end{lemma}
\begin{proof}
\begin{enumerate}
\item Under the assumption (\ref{e28}), by c), we conclude
that  $R_{c_1}$ is only obtained
at $(j,k,l)=(1,0,1)$, i.e. $j_1=1$.
\item Under the assumption (\ref{e29}), there exists
  $j^* \ge 2$ such that $ {j^*}^2\alpha_1^2 \le x_{b_1}
< (j^*+1)^2 \alpha_1^2$. We note that
\begin{eqnarray*}
(j^*+1)^2 \alpha_1^2 \begin{cases}
 < 2 {j^*}^2\alpha_1^2 < 2 x_{b_1}<\alpha_2^2  \qquad
 \qquad \text{if} \qquad j^* \ge 3,\\
 =9 \alpha_1^2 < \frac{9}{5} x_{b_1} < 2x_{b_1}
<\alpha_2^2 \qquad \text{if}
 \qquad  j^*=2.
\end{cases}
\end{eqnarray*}
\end{enumerate}
Hence, by b) and c), we conclude that
$$
R_{c_1}=\min \{f_{b_1}({j^*}^2 \alpha_1^2),f_{b_1}((j^*+1)^2\alpha_1^2)\},
$$
i.e., $j_1=j^*$ or $j_1=j^*+1$.
Note that, by b) and c), generically
$f_{b_1}({j^*}^2 \alpha_1^2)\ne f_{b_1}((j^*+1)^2\alpha_1^2)$.
The proof is complete.
\end{proof}

\begin{lemma}
\label{le5}
\begin{enumerate}
\item Condition (\ref{c7}) holds true under the assumption (\ref{e30}).
\item Generically, Condition (\ref{c7}) holds true under the assumption
      (\ref{e31}).
\end{enumerate}
\end{lemma}
\begin{proof}
Consider
$$
R_{c_2}= \min_{(j,k,1) \in \Lambda_1}\{2 (\sigma +1)f_{b_2}(\alpha_{jk}^2)\}.
$$
The rest part of the proof is the same as the proof of Lemma~\ref{le4}.
\end{proof}

\begin{lemma}
\label{le6}
 Assume (\ref{c7}),  $R \approx R_{c_2}$ and
 $Ro^2$ satisfies (\ref{e21}) for $(j,k,l)=(j_2, 0, 1)$, i.e.,
$Ro^2<\frac{(1-\sigma)\pi^2}{\sigma^2(1+\sigma) \gamma_{j_2 0 1}^6}
 $, then  $\{\beta_{j_2 0 11}(R),\beta_{j_2 0 12}(R)\}$
   ($\beta_{j_2 0 11}(R)=\bar{\beta}_{j_2 0 12}(R)$)
      is the only simple pair of complex eigenvalues
      of the problem (\ref{e1}) in space $\tilde{H}_1$
      satisfying
  \begin{equation}
\label{e32}
  Re(\beta_{j_2 0 11}(R)) \begin{cases}
<0 \,\,\,\, \text{if}\,\,\,\, R < R_{c_2},\\
=0 \,\,\,\, \text{if}\,\,\,\, R = R_{c_2},\\
>0 \,\,\,\, \text{if}\,\,\,\, R > R_{c_2},
                   \end{cases}
 \end{equation}
\begin{equation}
 Re \beta_{jklq}(R) < 0 \,\, \text{for} \,\,
   (\alpha_{jk}^2, l) \ne (\alpha_{j_2 0}^2, 1), \,q=1,2,3, \,
   R \,\, \text{near} \,\, R_{c_2}.
  \end{equation}
\end{lemma}
\begin{proof}
We only need to prove (\ref{e32}). Under the assumptions of the
lemma together with (\ref{e11}), (\ref{e19}) and (\ref{e20}),
 by the discussion in Case (2)
at the beginning of this subsection,
 we know that
$\{ \beta_{j_2 0 11}(R) , \beta_{j_2 0 12}(R) \}$
is the only simple pair of complex eigenvalues of
$L_R \mid _{\tilde{H}_1}$ with
$Re( \beta_{j_2 0 11}(R_{c_2}))=Re( \beta_{j_2 0 12}(R_{c_2}))=0$.
Since  $\beta_{j_2 0 13}(R)$ (real),  $\beta_{j_2 0 11}(R)$
and $\beta_{j_2 0 12}(R)$ are zeros of $f_{j_2 0 1}$, we know that
$$
\beta_{j_2 0 13}(R)=-(Re( \beta_{j_2 0 11}(R))+
 Re( \beta_{j_2 0 12}(R)))-(2\sigma+1)\gamma_{j_2 0 1}^2.
$$
Hence (\ref{e32}) is equivalent to
\begin{equation}
\label{e33}
\beta_{j_2 0 13}(R)\begin{cases}
>-(2\sigma+1)\gamma_{j_2 0 1}^2 \,\,\,\, \text{if}\,\,\,\, R < R_{c_2},\\
=-(2\sigma+1)\gamma_{j_2 0 1}^2 \,\,\,\, \text{if}\,\,\,\, R = R_{c_2},\\
<-(2\sigma+1)\gamma_{j_2 0 1}^2 \,\,\,\, \text{if}\,\,\,\, R >
R_{c_2},
                   \end{cases}
\end{equation}
which is true as shown in Figure~\ref{fig4}. This completes the proof.

\begin{figure}
 \centering \includegraphics[height=.5\hsize]{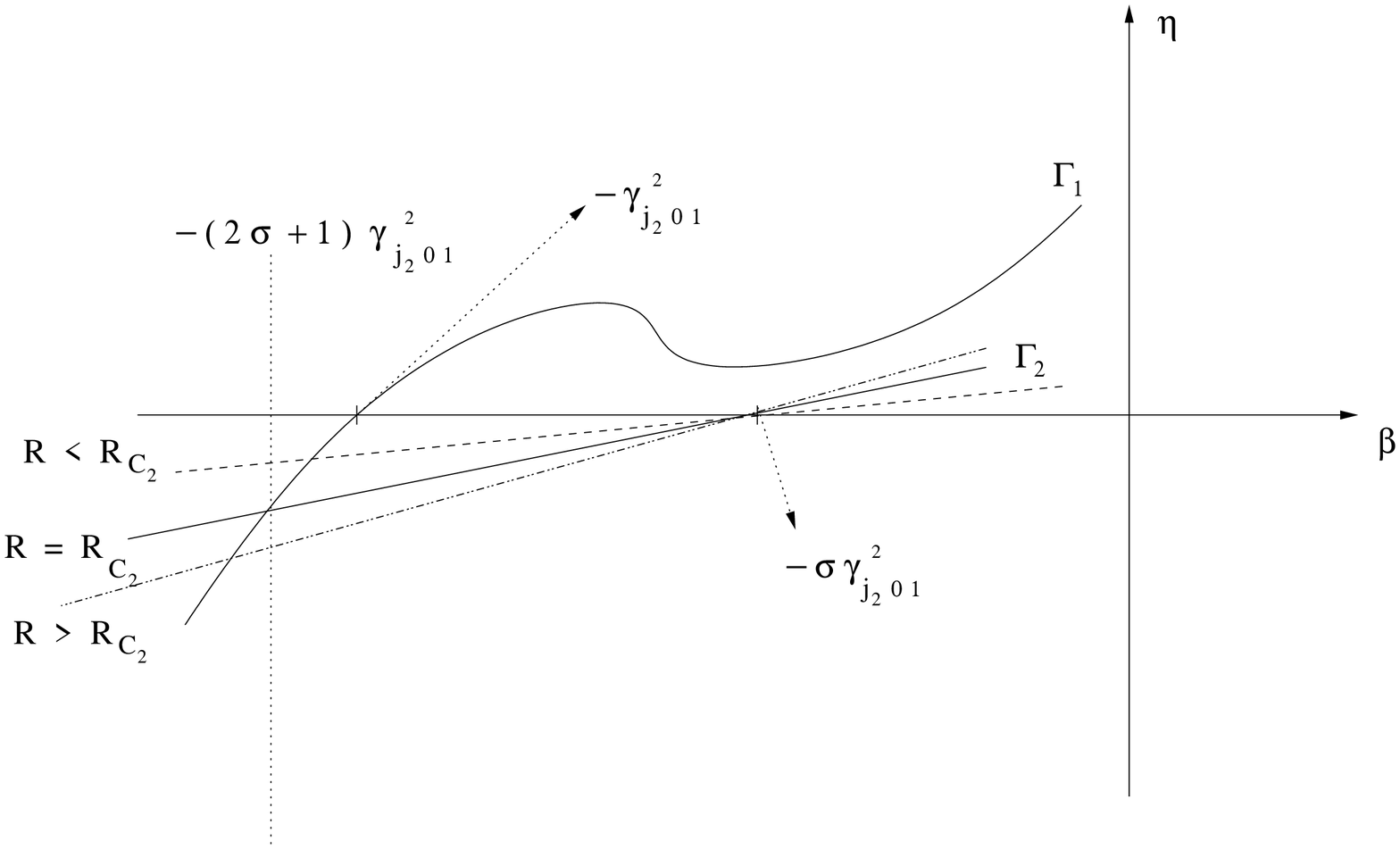}
 \caption{}
\label{fig4}
\end{figure}
\end{proof}

\begin{lemma}
\label{le8}
For fixed $\alpha_1$, $\alpha_2 >0$ and $\sigma > 1$,
$R_{c_1} \rightarrow \infty$ as $Ro \rightarrow 0$.
More precisely, $R_{c_1}= O(Ro^{-\frac{4}{3}})$.
\end{lemma}
\begin{proof}
Since $b_1=\frac{\pi^2}{\sigma^2 Ro^2}$,
by (\ref{e27}), $x_{b_1}=O(b_1^{\frac{1}{3}})$ as $Ro \rightarrow 0$.
Hence,
\begin{eqnarray*}
 R_{c_1}=O(f_{b_1}(x_{b_1}))=O(b_{1}^{\frac{2}{3}})=O(Ro^{-\frac{4}{3}}).
\end{eqnarray*}
\end{proof}

\section{Proof of Main Theorems}
\subsection{Center manifold reduction}
We are now in a position to reduce equations of (\ref{b1})-(\ref{b4})
to the center manifold.  For any $\psi = (U,T) \in H_1$, we have
\begin{align*}
\psi =& \sum_{(j,k,l)\in \Lambda_1}^{\infty} \sum_{q=1}^{3}
      (x_{jklq}\psi^{\beta_{jklq}}_1 + y_{jklq}\psi^{\beta_{jklq}}_2)\\
      & + \sum_{(j,k,0)\in \Lambda_2}(x_{jk0}\psi^{\beta_{jk0}}_1
            +y_{jk0}\psi^{\beta_{jk0}}_2)
         +\sum_{l=1}^{\infty}\sum_{q=1}^{3}x_{00lq}\psi^{\beta_{00lq}} .
\end{align*}
Under the assumption (\ref{c6}), the first
critical Rayleigh number is given by
\be
\label{f1}
R_{c_1} =  \frac{\gamma_{j_1 0 1}^6}{\alpha_{j_1 0}^2}
+ \frac{\pi^2}{\sigma^2 Ro^2 \alpha_{j_1 0}^2}.
\ee
 In this case, the multiplicity
of the first eigenvalue is two and the reduced
equations of  (\ref{b1})-(\ref{b4}) are given
by
\begin{equation}
\label{f2}
\left\{
\begin{aligned}
& \frac{d x_{j_1011}}{dt}=\beta_{j_1011}(R)x_{j_1011} +\frac{1}
  {<\psi_{1}^{\beta_{j_1011}},\Psi_{1}^{\beta_{j_1011}}>_H}
  <G( \psi, \psi),\Psi_{1}^{\beta_{j_1011}})>_H,\\
& \frac{dy_{j_1011}}{dt}=\beta_{j_1011}(R)y_{j_1011}+\frac{1}
  {<\psi_{2}^{\beta_{j_1011}},\Psi_{2}^{\beta_{j_1011}}>_H}
  <G( \psi, \psi),\Psi_{2}^{\beta_{j_1011}})>_H.\\
\end{aligned}
\right.
\end{equation}

Here for $\psi_1=(U_1,T_1)$, $\psi_2=(U_2,T_2)$ and
$\psi_3=(U_3,T_3)$,
\begin{eqnarray*}
G(\psi_1,\psi_2)=-( P(U_1\cdot\nabla)U_2,
                   (U_1\cdot\nabla)T_2 )^t
\end{eqnarray*}
and
\begin{align*}
<G(\psi_1,\psi_2),\psi_3>_H=-\int_{0}^{1}
    \int_{0}^{2\pi/\alpha_2}
    \int_{0}^{2\pi/\alpha_1} & [<(U_1\cdot\nabla)U_2,
         U_3>_{\mathbb R^3}\\
   & +(U_1\cdot\nabla)T_2 T_3
        ]dxdydz,
\end{align*}
where P is the Leray projection to
$L^2$ fields.
Let the center manifold function be denoted by
\begin{equation}
\label{f3}
\Phi=\sum_{\beta \ne \beta_{j_1011}}( \Phi_{1}^{\beta}(x_{j_1011},
      y_{j_1011})\psi_{1}^{\beta}+
     \Phi_{2}^{\beta}(x_{j_1011},y_{j_1011})\psi_{2}^{\beta}).
\end{equation}

The direct calculation shows that
\begin{equation}
\label{f4}
\begin{aligned}
& G(\psi^{\beta_{j_1011}}_1,\psi^{\beta_{j_1011}}_1)
  = -(0, \frac{A_1 \pi^2}{2 j_1 \alpha_1} \sin 2 j_1 \alpha_1 x,
     0, \frac{A_2 \pi}{2} \sin 2 \pi z )^t,\\
& G(\psi^{\beta_{j_1011}}_1,\psi^{\beta_{j_1011}}_2)
  =-( \frac{\pi^2}{2 j_1 \alpha_1}\cos 2 \pi z,
    \frac{A_1 \pi^2}{2  j_1 \alpha_1}
 (\cos 2 \pi z -\cos 2j_1 \alpha_1 x),0,0)^t,\\
& G(\psi^{\beta_{j_1011}}_2,\psi^{\beta_{j_1011}}_1)
  = -( \frac{-\pi^2}{2 j_1 \alpha_1} \cos 2 \pi z,
      \frac{-A_1 \pi^2}{ 2j_1 \alpha_1}(\cos 2 j_1 \alpha_1 x
        + \cos 2 \pi z ), 0, 0)^t,\\
& G(\psi^{\beta_{j_1011}}_2,\psi^{\beta_{j_1011}}_2)
  =-(0, \frac{- A_1 \pi^2 }{2 j_1 \alpha_1} \sin 2 j_1 \alpha_1 x,
    0, \frac{A_2 \pi}{2} \sin 2 \pi z)^t.
  \end{aligned}
\end{equation}

\begin{equation}
\label{f5}
\begin{aligned}
& G(\psi^{\beta_{j_1011}}_1,\Psi^{\beta_{j_1011}}_1)
  = -(0, \frac{C_1 \pi^2}{2 j_1 \alpha_1} \sin 2 j_1 \alpha_1 x,
     0, \frac{C_2 \pi}{2} \sin 2 \pi z )^t,\\
& G(\psi^{\beta_{j_1011}}_1,\Psi^{\beta_{j_1011}}_2)
  =-( \frac{\pi^2}{2 j_1 \alpha_1}\cos 2 \pi z,
    \frac{C_1 \pi^2}{2  j_1 \alpha_1}
 (\cos 2 \pi z -\cos 2j_1 \alpha_1 x),0,0)^t,\\
& G(\psi^{\beta_{j_1011}}_2,\Psi^{\beta_{j_1011}}_1)
  = -( \frac{-\pi^2}{2 j_1 \alpha_1} \cos 2 \pi z,
      \frac{-C_1 \pi^2}{ 2j_1 \alpha_1}(\cos 2 j_1 \alpha_1 x
        + \cos 2 \pi z ), 0, 0)^t,\\
& G(\psi^{\beta_{j_1011}}_2,\Psi^{\beta_{j_1011}}_2)
  =-(0, \frac{- C_1 \pi^2 }{2 j_1 \alpha_1} \sin 2 j_1 \alpha_1 x,
    0, \frac{C_2 \pi}{2} \sin 2 \pi z)^t,
  \end{aligned}
\end{equation}
where $A_1=A_1(\beta_{j_1011})$,  $A_2=A_2(\beta_{j_1011})$
       $C_1=C_1(\beta_{j_1011})$ and $C_2=C_2(\beta_{j_1011})$.

Hereafter, we make the following convention:
\begin{align*}
 & o(2)= o(x_{j_1011}^2+y_{j_1011}^2)+ O (\mid\beta_{j_1011}(R)\mid
  \cdot (x_{j_1011}^2+y_{j_1011}^2)),\\
 & o(3)= o((x_{j_1011}^2+y_{j_1011}^2)^{3/2})+O (\mid\beta_{j_1011}(R)\mid
  \cdot (x_{j_1011}^2+y_{j_1011}^2)^{3/2}),\\
 & o(4)= o((x_{j_1011}^2+y_{j_1011}^2)^{2})+O (\mid\beta_{j_1011}(R)\mid
  \cdot (x_{j_1011}^2+y_{j_1011}^2)^{2}).
\end{align*}
 By Theorem~\ref{thd2} and (\ref{f4})-(\ref{f5}), we obtain
\be
\Phi=\Phi_{1}^{\beta_{(2j_1) 00}}\psi_{1}^{\beta_{(2j_1)00}}+
     \Phi_{2}^{\beta_{(2j_1) 00}}\psi_{2}^{\beta_{(2j_1)00}}+
      \Phi_{1}^{\beta_{0021}}\psi_{1}^{\beta_{0021}}+o(2),
\ee
where
\begin{align*}
& \Phi_{1}^{\beta_{(2j_1)00}}=\frac{A_1 \pi^2}
 {\sigma \alpha_{(2j_1)0}^4} (x_{j_1011}^2 - y_{j_1011}^2)+o(2),
&&\psi_{1}^{\beta_{(2j_1)00}}=(0,
 -2j_1 \alpha_1 \sin 2 j_1 \alpha_1 x, 0, 0)^t,\\
& \Phi_{2}^{\beta_{(2j_1)00}}=\frac{A_1 \pi^2}
 { \sigma \alpha_{(2j_1)0}^4}( 2 x_{1011} y_{1011})+o(2),
&& \psi_{2}^{\beta_{(2j_1)00}}=(0,
  2 j_1 \alpha_1 \cos 2 \alpha_1 x, 0, 0)^t,\\
& \Phi_1^{\beta_{0021}}=\frac{-A_2}{8 \pi}
  (x_{j_1011}^2+y_{j_1011}^2)+o(2),
&& \psi_{1}^{\beta_{0021}}=(0,0,0, \sin 2 \pi z)^t.
\end{align*}

Note that for any $\psi_i \in H_1$($i=$ 1, 2, 3),
\begin{align}
\label{f7}
& <G(\psi_1,\psi_2),\psi_2>_H=0,\\
\label{f8}
& <G(\psi_1,\psi_2),\psi_3>_H=-<G(\psi_1,\psi_3),\psi_2>_H;
\end{align}
and for any $\psi_{i}\in E_{jkl}$ $(i=1,2,3)$,
\begin{equation}
\label{f9}
 <G(\psi_1,\psi_2),\psi_3>_H=0.
\end{equation}
The direct calculation shows that
\begin{equation}
\label{f10}
G(\tilde{\psi}, \psi^{\beta_{j_1 0 1 1}}_i)=0 \qquad \text{for} \qquad
\tilde{\psi} \in \{\psi_{1}^{\beta_{(2j_1)00}},
\psi_{2}^{\beta_{(2j_1)00}}, \psi_{1}^{\beta_{0021}}\} ,\, i=1,2.
\end{equation}

Then by $\psi = x_{j_1011} \psi^{\beta_{j_1011}}_1
     + y_{j_1011} \psi^{\beta_{j_1011}}_2
     + \Phi(x_{j_1011},y_{j_1011})$ and (6.4)-(6.10), we derive that
 \begin{align}
<G(\psi, \psi),& \Psi^{\beta_{j_1011}}_1>_H
\nonumber
\\
=   & <G(\psi^{\beta_{j_1011}}_1, \Phi), \Psi^{\beta_{j_1011}}_1>_H x_{j_1011}
+<G(\psi^{\beta_{j_1011}}_2, \Phi), \Psi^{\beta_{j_1011}}_1>_H y_{j_1011}+o(3),
\nonumber\\
=&-  <G(\psi^{\beta_{j_1011}}_1, \Psi^{\beta_{j_1011}}_1), \Phi>_H x_{j_1011}
- <G(\psi^{\beta_{j_1011}}_2,
 \Psi^{\beta_{j_1011}}_1),\Phi>_H y_{j_1011}+o(3),
\nonumber\\
= & -\frac{2 A_1 C_1 \pi^6}
 { \sigma \alpha_1 \alpha_2 \sigma_{(2j_1)0}^4}
  (x_{j_1011}^2-y_{j_1011}^2) x_{j_1011}
 -\frac{A_2 C_2 \pi^2}{ 8 \alpha_1 \alpha_2}
  (x_{j_1011}^2+y_{j_1011}^2) x_{j_1011} +o(3),
\nonumber \\
& -\frac{2 A_1 C_1 \pi^6}
 {  \sigma \alpha_1 \alpha_2 \alpha_{(2j_1)0}^4}
 (2x_{j_1011} y_{j_1011})y_{j_1011}
\nonumber\\
=&  -(\frac{2 A_1 C_1 \pi^6}
 {  \sigma \alpha_1 \alpha_2 \alpha_{(2j_1)0}^4}
     +\frac{A_2 C_2 \pi^2}{ 8 \alpha_1 \alpha_2} )
(x_{j_1011}^2+y_{j_1011}^2) x_{j_1011}   +o(3).
\nonumber
\end{align}
Similarly, we obtain
\begin{align*}
<G(\psi, \psi), \Psi^{\beta_{j_1011}}_2>_H =
 -(\frac{2 A_1 C_1 \pi^6}
 {  \sigma \alpha_1 \alpha_2 \alpha_{(2j_1)0}^4}
     +\frac{A_2 C_2 \pi^2}{ 8 \alpha_1 \alpha_2} )
(x_{j_1011}^2+y_{j_1011}^2) y_{j_1011}   +o(3).
\end{align*}

Hence, the reduction equations are given by
\be
\label{f11}
\left\{
\begin{aligned}
& \frac{dx_{j_1011}}{dt}=\beta_{j_1011}(R) x_{j_1011}
 + \delta (x_{j_1011}^2+y_{j_1011}^2) x_{j_1011}+o(3),\\
& \frac{dy_{j_1011}}{dt}=\beta_{j_1011}(R) y_{j_1011} + \delta
(x_{j_1011}^2+y_{j_1011}^2) y_{j_1011}+o(3),
\end{aligned}
\right.
\ee
where
\be
\label{f12}
\delta= - (\frac{2 A_1 C_1 \pi^4}{ \sigma \alpha_{(2j_1)0}^4}
           + \frac{A_2 C_2}{8})/ (\frac{ \pi^2}{ j_1^2 \alpha_1^2}
           (1+A_1 C_1) +1 + A_2 C_2) < 0.
\ee
A standard energy estimate on (\ref{f11}) together
with  the center manifold theory show that, for $R \le R_{c_1}$,
$(U,T)=0$ is locally asymptotically stable
for the problem (\ref{b1})-(\ref{b4}). Hence by Theorem~\ref{thd1},
the solutions to (\ref{b1})-(\ref{b4}) bifurcate from
$(U,T,R)=(0,R_{c_1})$ to an attractor $\Sigma_{R}$.
Moreover, by (\ref{f11})-(\ref{f12}) together with
Theorem 5.10 in \cite{amsbook}, we conclude that
 $\Sigma_{R}$ is homeomorphic to $S^{1}$ in $H$.

\subsection{Completion of the proof of Theorem 3.4}
In this subsection,
we prove that $\Sigma_{R}$ consists of steady state solutions.
It is clear that the first eigenvalue of $L_{R}|_{\tilde{H}_1}$ is simple
  for $R\approx R_{c_1}$.
By the Kransnoselski  bifurcation theorem (see among others Chow and Hale
\cite{ch} and  Nirenberg \cite{nirenberg}),
when $R$ crosses $R_{c_1}$,
the equations bifurcate from
the basic solution to a steady state solution in $\tilde{H}$.
Therefore the attractor $\Sigma_R$
contains at least one steady state solution.
Secondly, it's easy to check that the equations (\ref{b1})-(\ref{b4})
defined in $H$ are
translation invariant in the $x$-direction. Hence if
 $\psi_0(x,y,z)=(U(x,y,z),T(x,y,z))$ is a steady state solution,
 then $\psi_0(x+\rho,y,z)$ are steady state solutions as well. By the periodic
 condition in the $x$-direction, the set
\begin{align*}
S_{\psi_0}=\{\psi_0(x+\rho,y,z) | \rho \in \mathbb R \}
\end{align*}
is a cycle  homeomorphic to $S^1$ in $H$. Therefore the steady state of
 (\ref{b1})-(\ref{b4}) generates a cycle of steady state solutions.
Hence the bifurcated attractor $\Sigma_R$ consists of steady state
solutions. The proof of Theorem~\ref{thc4} is complete.

\subsection{Proof of Theorem 3.5}
The proof  follows directly from the classical Hopf bifurcation theorem and
Lemma~\ref{le6}.


\begin{thebibliography}{10}

\bibitem{CI}
{\sc P.~Cessi and G.~R. Ierley}, {\em Symmetry-breaking multiple equilibria in
  quasi-geostrophic, wind-driven flows}, J. Phys. Oceanogr., 25 (1995),
  pp.~1196--1202.

\bibitem{CD}
{\sc J.~Charney and J.~DeVore}, {\em Multiple flow equilibria in the atmosphere
  and blocking}, J. Atmos. Sci., 36 (1979), pp.~1205--1216.

\bibitem{cgsw}
{\sc Z.-M. Chen, M.~Ghil, E.~Simonnet, and S.~Wang}, {\em Hopf bifurcation in
  quasi-geostrophic channel flow}, SIAM J. Appl. Math., 64 (2003), pp.~343--368
  (electronic).

\bibitem{ch}
{\sc S.~N. Chow and J.~K. Hale}, {\em Methods of bifurcation theory}, vol.~251
  of Grundlehren der Mathematischen Wissenschaften [Fundamental Principles of
  Mathematical Science], Springer-Verlag, New York, 1982.

\bibitem{D}
{\sc H.~A. Dijkstra}, {\em Nonlinear Physical Oceanography: A Dynamical Systems
  Approach to the Large-Scale Ocean Circulation and El Ni\~no}, Kluwer Acad.
  Publishers, Dordrecht/ Norwell, Mass., 2000.

\bibitem{GC}
{\sc M.~Ghil and S.~Childress}, {\em Topics in Geophysical Fluid Dynamics:
  Atmospheric Dynamics, Dynamo Theory, and Climate Dynamics}, Springer-Verlag,
  New York, 1987.

\bibitem{henry}
{\sc D.~Henry}, {\em Geometric theory of semilinear parabolic equations},
  vol.~840 of Lecture Notes in Mathematics, Springer-Verlag, Berlin, 1981.

\bibitem{IS}
{\sc G.~Ierley and V.~A. Sheremet}, {\em Multiple solutions and
  advection-dominated flows in the wind-driven circulation. i: Slip}, J. Marine
  Res., 53 (1995), pp.~703--737.

\bibitem{JJG}
{\sc S.~Jiang, F.-F. Jin, and M.~Ghil}, {\em Multiple equilibria, periodic, and
  aperiodic solutions in a wind-driven, double-gyre, shallow-water model}, J.
  Phys. Oceanogr., 25 (1995), pp.~764--786.

\bibitem{JG}
{\sc F.~F. Jin and M.~Ghil}, {\em Intraseasonal oscillations in the
  extratropics: Hopf bifurcation and topographic instabilities}, J. Atmos.
  Sci., 47 (1990), pp.~3007--3022.

\bibitem{LG}
{\sc B.~Legras and M.~Ghil}, {\em Persistent anomalies, blocking and variations
  in atmospheric predictability}, J. Atmos. Sci., 42 (1985), pp.~433--471.

\bibitem{LTWa}
{\sc J.~L. Lions, R.~Temam, and S.~Wang}, {\em New formulations of the
  primitive equations of the atmosphere and applications}, Nonlinearity, 5
  (1992), pp.~237--288.

\bibitem{LTWb}
\leavevmode\vrule height 2pt depth -1.6pt width 23pt, {\em On the equations of
  large-scale ocean}, Nonlinearity, 5 (1992), pp.~1007--1053.

\bibitem{La}
{\sc E.~N. Lorenz}, {\em Deterministic nonperiodic flow}, J. Atmos. Sci., 20
  (1963), pp.~130--141.

\bibitem{Lb}
\leavevmode\vrule height 2pt depth -1.6pt width 23pt, {\em The mechanics of
  vacillation}, J. Atmos. Sci., 20 (1963), pp.~448--464.

\bibitem{b-book}
{\sc T.~Ma and S.~Wang}, {\em Bifurcation Theory and Applications}, vol.~53 of
  World Scientific Series on Nonlinear Science, Series A, World Scientific,
  2005.

\bibitem{mw-db1}
\leavevmode\vrule height 2pt depth -1.6pt width 23pt, {\em Dynamic bifurcation
  of nonlinear evolution equations and applications}, Chinese Annals of
  Mathematics, 26:2 (2005), pp.~185--206.

\bibitem{amsbook}
\leavevmode\vrule height 2pt depth -1.6pt width 23pt, {\em Geometric Theory of
  Incompressible Flows with Applications to Fluid Dynamics}, vol.~119 of
  Mathematical Surveys and Monographs, American Mathematical Society,
  Providence, RI, 2005.

\bibitem{MB}
{\sc S.~Meacham and P.~Berloff}, {\em Instability of a steady, barotropic, wind
  driven circulation}, J. Mar. Res., 55 (1997), pp.~885--913.

\bibitem{BM}
\leavevmode\vrule height 2pt depth -1.6pt width 23pt, {\em On the stability of
  the wind-driven circulation}, J. Mar. Res., 56 (1998), pp.~937--993.

\bibitem{nirenberg}
{\sc L.~Nirenberg}, {\em Topics in nonlinear functional analysis}, Courant
  Institute of Mathematical Sciences New York University, New York, 1974.
\newblock With a chapter by E. Zehnder, Notes by R. A. Artino, Lecture Notes,
  1973--1974.

\bibitem{Pa}
{\sc J.~Pedlosky}, {\em Resonant topographic waves in barotropic and baroclinic
  flows}, J. Atmos. Sci., 38 (1981), pp.~2626--2641.

\bibitem{Pb}
\leavevmode\vrule height 2pt depth -1.6pt width 23pt, {\em Geophysical Fluid
  Dynamics, 2nd Edition}, Springer-Verlag, New-York, 1987.

\bibitem{SDG}
{\sc S.~Speich, H.~Dijkstra, and M.~Ghil}, {\em Successive bifurcations in a
  shallow-water model, applied to the wind-driven ocean circulation}, Nonlin.
  Proc. Geophys., 2 (1995), pp.~241--268.

\bibitem{S}
{\sc H.~Stommel}, {\em Thermohaline convection with two stable regimes of
  flow}, Tellus, 13 (1961), pp.~224--230.

\bibitem{temam}
{\sc R.~Temam}, {\em Infinite-dimensional dynamical systems in mechanics and
  physics}, vol.~68 of Applied Mathematical Sciences, Springer-Verlag, New
  York, second~ed., 1997.

\bibitem{Va}
{\sc G.~Veronis}, {\em An analysis of wind-driven ocean circulation with a
  limited fourier components}, J. Atmos. Sci., 20 (1963), pp.~577--593.

\bibitem{Vb}
\leavevmode\vrule height 2pt depth -1.6pt width 23pt, {\em Wind-driven ocean
  circulation, part ii: Numerical solution of the nonlinear problem}, Deep-Sea
  Res., 13 (1966), pp.~31--55.

\end{thebibliography}

\end{document}